\documentclass[aps,prl,twocolumn]{revtex4}
\usepackage[usenames]{color}
\newcommand{\comment}[1]{}
\usepackage{graphicx}
\usepackage{amssymb}
\usepackage{epsfig}
\newcommand{\bra}[1]{\langle {#1} |}
\newcommand{\ket}[1]{| {#1} \rangle}

\newcommand{\ketn}[1]{ {#1} \rangle}

\begin{document}

\bibliographystyle{revtex}
\title{Classifying Vortices in $S=3$ Bose-Einstein Condensates}
\author{Ryan Barnett$^{1}$, Ari Turner$^2$, and 
Eugene Demler$^2$}
\affiliation{$^1$Department of Physics, California Institute of Technology, MC 114-36, Pasadena, California 91125,USA}
\affiliation{$^2$Department of Physics, Harvard University, Cambridge, Massachusetts 02138, USA}
\date{\today}
\begin{abstract}
Motivated by the recent realization of a $^{52}$Cr Bose-Einstein condensate,
we consider the phase diagram of a general spin-three condensate as a 
function of its scattering lengths.  We classify each phase 
according to its reciprocal spinor,
using a method developed in a previous work.
We show that such a classification can be naturally extended to
describe the vortices for a spinor condensate 
by using the topological theory of defects.
To illustrate, we systematically describe the types of vortex 
excitations for each phase of the spin-three condensate.
\end{abstract}

\maketitle

Over the past few years, there has been remarkable experimental progress
in the study of spinor Bose-Einstein condensates 
\cite{stenger98,schmaljohann04,chang04,griesmaier05,sadler06}.
One of the most intriguing examples is the recent realization
of a  $^{52}$Cr \cite{griesmaier05} condensate 
which has since received substantial theoretical attention
\cite{santos06,diener06,makela06b,bernier06}.  Atomic chromium
has six electrons in its outermost shell which combine into a
state with maximal spin $S=3$
according to Hund's rules.  These electrons also give rise to a
magnetic dipole moment which is much larger than that of the
alkalis where dipolar effects are typically unimportant.  
In fact, such dipolar effects have been observed in the
expansion profiles of $^{52}$Cr \cite{stuhler05}.  Dipolar effects
are important for magnetic traps where the spin degrees are
frozen out.  However, when purely optical traps are used it has been
argued that the spin exchange interaction will 
overwhelm the dipolar interaction \cite{diener06}.
Due to such spin interactions, 
a dilute gas of spin three bosons will interact 
according to the pair potential
\begin{eqnarray}
\label{pair-potential}
V_{\rm p}({\bf r_1}-{\bf r_2})=\delta({\bf r_1}-{\bf r_2})
(g_0 {\cal P}_0 + g_2 {\cal P}_2 + g_{4}{\cal P}_2 +  g_{6} {\cal P}_{6})
\label{Vcontact}
\end{eqnarray}
where ${\cal P}_S$ projects into the state with
total spin $S$ and $g_S=4\pi \hbar^2 a_S/m$ where $a_S$ is the scattering
length corresponding to spin $S$.  For the spin-three problem, this
potential gives rise to a phase diagram (as a function of
the possible values of the scattering lengths) of great richness 
as was revealed in \cite{diener06,santos06}.  
In addition to having such rich phase diagrams,  spinor condenstates 
in general and the $^{52}$Cr 
condensate in particular will have intriguing
topological excitations.  Topological excitations for spin-one, spin
three-halves, 
and spin-two condenstates have received considerable attention
in the literature 
\cite{zhou01,stoof01,wu05,makela06a,mukerjee06,semenoff06} since 
the pioneering work of Ho \cite{ho98} and Ohmi and 
Machida \cite{ohmi98}.  Also, recently topological spin textures 
have been observed in $^{87}$Rb atoms in the spin-one hyperfine
state \cite{sadler06}.

In a previous publication \cite{barnett06}, we have
developed a way of classifying spinor BEC and Mott insulating 
states and illustrated
the method with an application to the spin-two boson problem.
More specifically, we showed that it is natural to identify 
a spin $S$ state $\ket{\psi}=\sum_{\alpha=-S}^{S} A_{\alpha} \ket{\alpha}$ 
with $2S$ points on the unit sphere $(\theta,\phi)$.  
In fact, there is a one-to-one correspondence between
a spinor and a ``reciprocal spinor'' up to phase and normalization
\begin{equation}
\lbrace A_{\alpha} \rbrace  \leftrightarrow \lbrace \zeta_{i} \rbrace
\end{equation}
where $\zeta=\tan(\theta/2)e^{i\phi}$ is the stereographic image on the complex
plane of the point $(\theta,\phi)$ on the unit sphere.
The set of such complex numbers $\lbrace \zeta_i \rbrace$ 
are explicitly constructed by finding the coherent states 
$\ket{\zeta}=\sum_{\alpha=0}^{2S} \sqrt{2S \choose \alpha}
\zeta^{\alpha} \ket{S-\alpha}$  
which are orthogonal to the spinor states
$\ket{\psi}$.  The
characteristic equation $f_{\psi}(\zeta)\equiv \bra{\psi}\ketn{\zeta}$
will then be a polynomial in $\zeta$ of order $2S$ and its 
$2S$ roots projected on the sphere will
immediately give the symmetries of the spinor.
Spin rotations which leave this set of points invariant will
take a spinor $A_{\alpha}$ into itself modulo an overall phase
which can be computed through geometrical considerations.
Thus this method gives a complete description of the spinor state.

\begin{figure}
\includegraphics[width=3.5in]{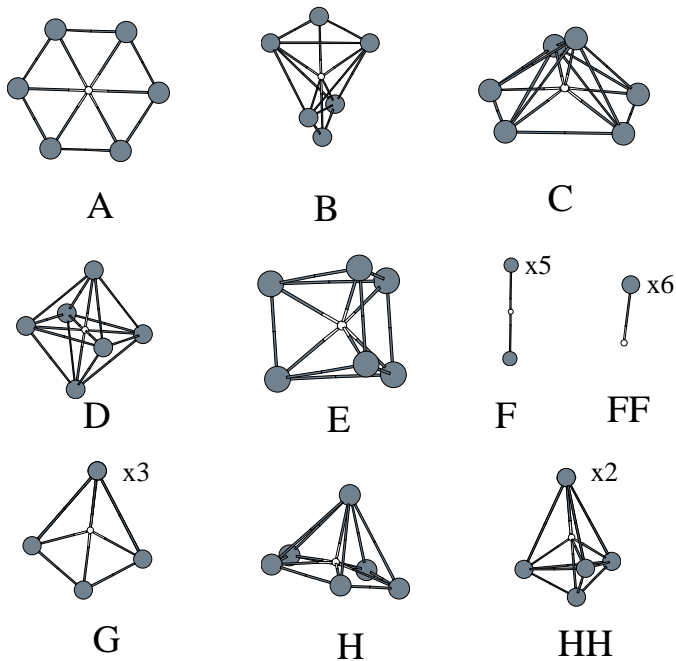}
\caption{Reciprocal spinors. Points on the unit sphere (gray spheres)
having the transformation properties
of the wave functions given in Table~\ref{table}.  
The small white spheres are
put at the origin as a visual aid.  Points which correspond to degenerate
roots are marked.}
\label{Fig:shapes}
\end{figure}

In this work, we show that such a method is also useful for
classifying
the spinor vortices.  
To illustrate, we focus on the spin-three problem, motivated by
the recent experimental development \cite{griesmaier05}. 
We first reproduce 
the phase diagrams for spin-three condensates obtained previously in 
\cite{diener06,santos06}.
Then we find the symmetries of each phase (see Fig.~\ref{Fig:shapes}) by 
applying the method developed in
\cite{barnett06}.  We then systematically describe the types of topological
excitations for such a system using the topological 
theory of defects \cite{mermin79}.   Such a scheme enumerates
all possible types of vortex excitations, 
though
it does not determine which are energetically stable against dividing into
multiple vortices.
Finally, we will briefly discuss the resulting microscopic wave functions
for the vortex excitations.  
Such understanding of vortices
will be relevant for rotating $^{52}$Cr condensates in optical traps.

In general, the topological vortex excitations (line defects)
of an ordered medium are given
by the first homotopy group $\pi_1 (R)$ (the group of closed loops) 
of the order parameter space  $R$.
More specifically, the conjugacy classes
of $\pi_1(R)$ correespond to the vortex types.
Moreover, the way such vortices combine
is determined by the multiplication table of the conjugacy classes.
For a pedagogical review of the topological theory of defects, 
see \cite{mermin79} (whose notation we will adopt).  
A theorem from algebraic topology 
(see e.g. \cite{hatcher02}) 
which is used as a reliable tool for computing
the first homotopy group is the following.  Let $G$ be a simply 
connected group and $H$ be a subgroup of $G$.  Then 
\begin{equation}
\label{theorem}
\pi_1(G/H)=H/H_0
\end{equation}
where $H_0$ are the elements of $H$ which are connected to the 
identity by continuous transformations.  To apply this theorem
to our system, we need to take $G$ to be a group of operators
that act transitively on the spin groundstates and 
$H$ to be the subgroup of $G$
which leave the particular spinor state we are considering invariant. Then
$R$ has the same topology as $G/H$, so the theorem applies.
The first guess would be to
use elements like 
$e^{i\alpha} e^{-i {\bf S} \cdot {\bf \hat{n}} \beta}$ (where
$\bf S$ is the spin-three matrix and $\bf \hat{n}$ is a unit
vector) which
act directly on the spinor states and superfluid phase.
However, these elements form a representation of $SO(3) \otimes U(1)$, 
which is not simply connected, so Eq.~(\ref{theorem}) 
will not apply.  Thus
we need instead to take the simply connected group 
$G= SU(2)\otimes T$ where $T$ is the group
of translations in 1d with elements satisfying the property $T(x)T(y)=T(x+y)$
(which will be simply connected).  This is the covering group
of $SO(3)\otimes U(1)$.
In this work we will use the 
representation of $SU(2)$ by Pauli matrices, thus we describe its elements in
the form
$e^{-i \frac{{\bf \sigma}\cdot \hat{\bf n}}{2}\Theta}$.
We then  use the homomorphism
$\varphi : SU(2)\otimes T \rightarrow  
SO(3)\otimes U(1)$ 
given by
$\varphi\left( T(x) e^{-i \frac{{\bf \sigma}\cdot {\bf \hat{n}}}{2} 
\Theta}\right)= e^{ix}e^{-i {\bf F}\cdot {\bf \hat{n}} \Theta}$ 
which ``lifts'' our group acting directly on the spinor states
into the larger simply connected group, $G$.  The subgroup $H \subseteq G$
will be the collection of elements $h \in G$ which leave  
the particular spinor under consideration
invariant.  Once $H$ is constructed
by such a method,
Eq.~(\ref{theorem}) can then be directly applied to construct 
the first homotopy group.

As we will see, the majority of the phases in the spin-three
phase diagram have only 
discrete symmetries.
For such states, $H_0$ is the
trivial group containing one element.  For this situation,
our computational theorem reduces to
\begin{equation}
\label{theorem_discrete}
\pi_1(G/H)=H.
\end{equation}
Therefore, constructing $\pi_1$ for such states becomes straightforward:
One first finds the rotations in $SO(3)$ which leave invariant the points 
on the unit sphere $\lbrace \zeta_\alpha \rbrace$ given by our 
classification scheme \cite{barnett06}.  
For this purpose, it is helpful to note the
possible finite subgroups of $SO(3)$.  They are the cyclic groups
$C_n$, the dihedral groups $D_n$, the tetrahedral group $T$,
the octahedral group $O$, and the icosahedral group $Y$ (see
Table~\ref{table}).
One then lifts this
subgroup of $SO(3)$ into the twice-as-large group $SU(2)$.  On top
of this ``spinor scaffolding''  
one can impose a superfluid winding $x$, but 
the allowed values of  $x$  are related to the spin rotation
chosen.
The winding is given in the form $x=-\lambda + 2\pi k$
where the integer $k$ is arbitrary.  
The offset $\lambda$
is fixed and can be determined geometrically through the
relation
\begin{equation}
\label{Eq:lambda}
\lambda=(r-S)\Theta
\end{equation}
where $\Theta$ is the angle of rotation, $r$ is the multiplicity of
the spin-root on the axis of rotation, and $S$ is the total spin
(see appendix for proof).
This information on the superfluid
phase determines the translation group elements $T(x)$.
This provides a direct way of
constructing $H$ and therefore classifying the possible vortices 
which we illustrate below for the spin-three problem.

Now we move on to discuss the possible phases of the spin-three
condensate in the absence of an external magnetic field.  
The interaction hamiltonian determined by the pair
potential Eq.~(\ref{pair-potential}) is given by
\begin{equation}
{\cal H}_{\rm int}=\frac{1}{2 L^3} \sum_{S,m} g_{S} a_{\alpha}^{\dagger}
a_{\beta}^{\dagger} \bra{\alpha \beta}\ketn{Sm} \bra{Sm} \ketn{\alpha'\beta'} 
a_{\alpha'} a_{\beta'}
\end{equation}
where $a_{\alpha}^{\dagger}$ creates a zero-momentum
boson in the spin $\alpha$
state, $\bra{\alpha \beta}\ketn{Sm}$ are Clebsch-Gordan coefficients, 
and the sums over the Greek indices are implicit.  Note that
this interaction energy alone is sufficient to determine the spinor
ground state since the rest of the hamiltonian will not have a spinor
preference.  Using the
variational state 
$\ket{\psi} = \frac{1}{\sqrt{N!}} 
\left( A_{\alpha} a_{\alpha}^{\dagger} \right)^{N} \ket{0}$,
where the coefficients are normalized $A^{*}_{\alpha}A_{\alpha}=1$
the interaction energy to leading order in $N$ can be found by
replacing $a_{\alpha}$ by the scalar $\sqrt{N} A_{\alpha}$
giving the interaction energy of
\begin{equation}
\label{varEnergy}
E_{\rm int}/N^2=\frac{1}{2L^3} \sum_{S} g_{S} P(S)
\end{equation}  
where
\begin{equation}
\label{Eq:P}
P(S)=\sum_{m} |A_{\alpha}^{*}A_{\beta}^{*} \bra{\alpha \beta}\ketn{Sm}|^2.
\end{equation}
We numerically minimize this energy as a function of the
seven complex variational parameters $A_{\alpha}$ under the constraint
of normalization $A^{*}_{\alpha}A_{\alpha}=1$ for all possible values
of the scattering lengths, and reproduce the phase diagram given
in \cite{santos06,diener06}. 
We give explicit forms of the
wave functions up to $SO(3)$ rotation which are summarized in
Table~\ref{table} where, for comparison, we use the same 
notation as was used in \cite{diener06}.  A subtle difference 
is that we find the additional 
state HH which is degenerate with the phase H reported in \cite{diener06}.
It is important to note that the unspecified angles 
$\eta$ and $\chi$ will take unique values for particular regions
in the phase diagram.  That is, there is no mean-field degeneracy
as was reported for the spin-two case \cite{barnett06}.

\begin{table*}
\begin{ruledtabular}
\begin{tabular}{c|c|c|c}
& $A_{\alpha}$ & $SO(3)$ subgroup & $N_{\rm
vort}$ \\ \hline A &
$\left(\frac{1}{\sqrt{2}},0,0,0,0,0,\frac{1}{\sqrt{2}}\right)$ &
$D_6$ & $9 \times {\mathbb Z}_v$ \\ \hline B &
$\left(0,\frac{\sin(\eta)}{\sqrt{2}},0,\cos(\eta),0,
-\frac{\sin(\eta)}{\sqrt{2}},0\right)$  & $D_2$ & $5\times
{\mathbb Z}_v$\\ \hline C & $\left(0,\frac{\sin(\chi)\cos(\eta)}{\sqrt{2}},
\frac{i \sin(\chi)\sin(\eta)}{\sqrt{2}},\cos(\chi), \frac{-i
\sin(\chi)\sin(\eta)}{\sqrt{2}},
\frac{\sin(\chi)\cos(\eta)}{\sqrt{2}},0\right)$  & $C_2$ &
$4\times{\mathbb Z}_v$ \\ \hline D &
$\left(0,\frac{1}{\sqrt{2}},0,0,0,\frac{1}{\sqrt{2}},0\right)$ &
$O$ & $8\times{\mathbb Z}_v$ \\ \hline E &
$\left(\frac{\sin(\eta)}{\sqrt{2}},0,0,\cos(\eta),0,0,
\frac{\sin(\eta)}{\sqrt{2}}\right)$ & $D_3$ & $6\times{\mathbb Z}_v$ \\
\hline F & $\left(0,1,0,0,0,0,0)\right)$  & $U(1)$ & $4$ \\
\hline FF & $\left(1,0,0,0,0,0,0\right)$  & $U(1)$ & $6$ \\
\hline G & $\left(0, \sin(\eta),0,\cos(\eta),0,0,0 \right)$ & $C_2$ & $4\times{\mathbb Z}_v$ \\ \hline H & $\left(
\sin(\eta),0,0,0,0,\cos(\eta),0\right)$ 
& $C_5$ & $10\times{\mathbb Z}_v$ \\ \hline HH &
$\left(0,\sin(\eta),0,0,\cos(\eta),0,0 \right)$ 
& $C_3$ & $6\times{\mathbb Z}_v$
\end{tabular}
\end{ruledtabular}
\caption{Wave functions  
$\ket{\psi} = \frac{1}{\sqrt{N!}} 
\left( A_{\alpha} a_{\alpha}^{\dagger} \right)^{N} \ket{0}$
for the possible spinor phases corresponding
to global minima of (\ref{varEnergy}) for particular scattering lengths
up to $SO(3)$ invariance. 
Also shown is  the subgroup of $SO(3)$ corresponding to the
symmetries of the wave function. The last column gives the
number of conjugacy classes of $\pi_1$ for each spin state, which
gives the number of possible vortex excitations. 
The ``$\times {\mathbb Z}_v$''
in this column 
reflects the fact that an arbitrary multiple of $2\pi$ can always be added to 
the winding in the superfluid phase.}
\label{table}
\end{table*}

We now apply the classification scheme developed in 
\cite{barnett06} to give the symmetries of the possible spinor
condensates, and the results are given in Fig.~\ref{Fig:shapes}.
Shown are reciprocal spin-three spinors which are six points
on the unit sphere which make the symmetries explicit.
All of the phases except for F and FF have no
continuous symmetries.  As was explained in \cite{diener06,santos06},
all of the scattering lengths for $^{52}$Cr are known except $a_0$.  This
gives a line in parameter space of possible ground states for such
a condensate where the phases A, B, and C are the
candidate ground states of the $^{52}$Cr condensate.

We will now discuss the
topological excitations of each phase.  For each case we will take 
$G=T\otimes SU(2)$ which will act on the spinor states via
the homomorphism $\varphi$.  The goal is to find the 
isotropy group $H$ which leaves the particular spinor state under
consideration invariant.
We start with the two states having continuous
symmetries, namely F and FF.  For state FF, 
taking the state to be oriented along
the $z$-axis, we find that the only operators leaving the spin
state invariant are 
$e^{i S \Theta}e^{-i S_z \Theta}$ with $S=3$.  
Note that the phase factor $e^{i S \Theta}$ is necessary
to cancel the phase from rotation.  Immediately, we find that 
this gives the isotropy group
\begin{eqnarray}
H&=&  
\left\lbrace T(S \Theta + 2\pi n) 
e^{-i \frac{\sigma_z }{2} \Theta}  : \Theta \in 
{\mathbb R} , n\in {\mathbb Z} \right\rbrace 
\\
H_0 &=& \left\lbrace T(S \Theta ) e^{-i \frac{\sigma_z }{2} \Theta}  : 
\Theta \in 
{\mathbb R} ,  \right\rbrace.
\end{eqnarray}
The first homotopy group is then computed to be 
$
\pi_1({\rm FF})=H/H_0 = {\mathbb Z}_{2S}
$
consistent with that in \cite{makela06a,mukerjee06}.
Carrying through the similar analysis for state F,
we find $\pi_1({\rm F})={\mathbb Z}_{2(S-1)}$.
Note that since these groups are abelian, the  
conjugacy classes are the original group elements 
giving.  The number of  conjugacy classes for these
cases
are $2S=6$ (FF) and $2(S-1)=4$ (F). 
The number of possible types of vortex excitations is
smaller by one here since  the identity corresponds 
to the absence of a vortex.
As noted before, this is quite a nontrivial result.  
This is in contrast
to the spinless case where there are infinitely many types of
vortices corresponding to different windings in the superfluid
phase as the vortex is circled (with $\pi_1={\mathbb Z}$).
The underlying reason for this is that rotating such a spinor
along its symmetry axis is identical to changing the superfluid
phase.

We now move on to consider the remaining phases which have only
discrete symmetries.  For these situations, as mentioned before,
$H_0$ is the trivial group, so we can apply Eq.~(\ref{theorem_discrete})
directly.  We will start with the phases having the smallest isotropy groups,
namely phases C and G.  For phase G, with the spinor wave function written
as in Table~\ref{table}, the only symmetry is a rotation by $\pi$
about the $z$-axis.  Such an operation will not change the
phase of the spinor, i.e. $e^{-i S_z \pi}\ket{\psi}=\ket{\psi}$.
We then find for the isotropy group
\begin{equation}
\pi_1({\rm G}) = H =\left\lbrace \pm  \tau_0,   \pm i \sigma_z \right\rbrace
\otimes \left\lbrace T(2\pi n) : n \in {\mathbb Z} \right\rbrace.
\end{equation}
where $\tau_0$ is the $2\times 2$ identity matrix and we 
note that $e^{\pm i\frac{\sigma_z}{2}\pi}=\pm i\sigma_z$.
This group is abelian, so the conjugacy classes are just the 
group elements themselves.  Note that the translational part
of $H$ merely tells us that an arbitrary winding in the superfluid
phase can be imposed on any vortex state.  \emph {In the following, we will
suppress the translational part of $H$}, keeping this in mind.
The phase C will have a similar isotropy group, but we note here
that rotation by $\pi$ about the symmetry axis will give a phase
of $e^{i\pi}=-1$.

Phases E, H, and HH have  $n$-fold rotational symmetries
about a particular axis.
Phases E and HH have a three-fold rotational symmetry about their symmetry
axes, and the first homotopy group is found to be
\begin{equation}
\pi_1({\rm E,HH})=\left\lbrace 
\pm e^{-i \frac{\sigma_z}{2}\frac{2\pi n}{3}}: n=1,2,3
\right\rbrace.
\end{equation}
On the other hand, 
phase H has a five-fold rotational
symmetry, and its first homotopy group is similarly
\begin{equation}
\pi_1({\rm H})=\left\lbrace 
\pm e^{-i \frac{\sigma_z}{2}\frac{2\pi n }{5}}: n=1,\ldots, 5
\right\rbrace
\end{equation}
As before, these groups are abelian.  
Therefore, the types of possible vortices
will correspond to rotating the polyhedra about the axes of symmetries
$n$ times as the vortex is circled.  In addition, the phase winds
by $-\lambda+2\pi k$ where $k$ is an arbitrary integer and 
$\lambda=0,-\frac{4\pi n}{5},-\frac{2\pi n}{3}$ for the phases 
E,H, and HH respectively. 

Phase B is the first phase we encounter that has a nonabelian
isotropy group.  This phase has a rotational symmetry of 
$\pi$ about the $z$-axis.  In addition, it also has the 
symmetry of a rotation about two axes in the $xy$ plane,
namely $(n_x,n_y)=\frac{1}{\sqrt{2}}(1,\pm 1)$.  The isotropy
group for this is
\begin{equation}
\pi_1({\rm B})=\left\lbrace 
\pm \tau_0, \pm i \sigma_z, \pm \frac{i}{\sqrt{2}} (\sigma_x \pm \sigma_y)
\right\rbrace
\end{equation}
Since this group is nonabelian, the conjugacy classes are not
just the original group.  This group is isomorphic to the
quaternion group having eight elements.
The conjugacy classes corresponding to each possible type
of vortex excitation are:
\begin{eqnarray}
&&
C_0 = \left \lbrace \tau_0 \right \rbrace, \;
\bar{C}_0 =\left \lbrace -\tau_0 \right \rbrace, \;
C_{z} = \left \lbrace \pm i \sigma_z \right \rbrace, \;
\\
&&
\nonumber
C_{xy}=
\left \lbrace \pm \frac{i}{\sqrt{2}}(\sigma_x+\sigma_y)\right \rbrace, \;
\\
&&
\nonumber
\bar{C}_{xy}=
\left \lbrace \pm \frac{i}{\sqrt{2}}(\sigma_x-\sigma_y)\right \rbrace.
\end{eqnarray}

Phases A and D have larger isotropy groups;  we will not write
out explicit representations here, but will merely state the results.
For phase A which has the symmetry
of the hexagon as seen in Fig.~\ref{Fig:shapes} there is a six-fold rotational
symmetry about the $z$-axis.  In addition to this there are six
axes in the $xy$ plane for which there will be a two-fold rotational
symmetry.  This will make $\pi_1({\rm A})$ have $24$ elements, which will
have $9$ conjugacy classes.  Phase D has the symmetry of the octahedron.
For this, $\pi_1({\rm C})$ will have $48$ elements and $8$ conjugacy
classes.  All of these results are summarized in Table~\ref{table}.

Now that the classification scheme for vortices based on homotopy
theory has been completed, we will  briefly 
consider the resulting microscopic wave functions.
We can express the order parameter explicitly
in terms of the spinor as 
$\psi_{\alpha}({\bf r})=\psi({\bf r}) A_\alpha ({\bf r})$ and
write $\psi({\bf r})=|\psi_0|e^{i\theta({\bf r})}$.
Note that the amplitude of $\psi({\bf r})$ is taken to have 
constant magnitude which is realistic provided we are sufficiently far
from the vortex core.  
For $A_{\alpha}({\bf r})$ it will prove
useful to use the spinor-ket notation and write 
$\ket{A({\bf r})}=e^{-i {\bf S}\cdot \hat{\bf n}\beta({\bf r})} \ket{A_0}$
where $\ket{A_0}$ is the reference state  which can be taken as
one of the spinors given in Table~\ref{table}.  In this equation,
$\hat{\bf n}$ is the axis about which the spinor is rotated as the vortex
is circled and $\theta({\bf r}),\beta({\bf r})$ give the spatial dependence
of the superfluid phase and spinor rotation around the vortex. These
functions must satisfy the requirement that the original wave function
must be recovered after the vortex is completely circled.  For instance,
suppose the spinor state under consideration is invarient under a 
rotation by angle $\Theta$.  Then a vortex at the origin
corresponding to this symmetry is given by 
$\theta({\bf r})=(-\frac{\lambda}{2 \pi} +k) \varphi$ 
and $\beta({\bf r})= \frac{\Theta}{2\pi}  \varphi$ where $\varphi$
is the azimuthal angle, $k$ is the winding number, and $\lambda$
is given by Eq.~(\ref{Eq:lambda}). 

This work was supported by the NSF grant DMR-0132874 and
the Harvard-MIT CUA. RB was supported by  
the Sherman Fairchild Foundation.
We thank 
R. Diener, 
T.-L. Ho, and
G. Refael
for useful discussions.

Note added: After this work was completed, we became aware of a work 
by S.-K. Yip \cite{yip06} 
which considers a similar problem of vortices in spinor BECs, 
but uses a different approach.

\section{Appendix: Proof of Eq.~(\ref{Eq:lambda})}

The geometrical reciprocal spinors determine the rotational symmetries of
the spinors. Associated with each such symmetry is a phase factor which we
will calculate here. Suppose the rotation through an angle $\Theta$ about 
the axis $\mathbf{\hat{n}}$ is a symmetry of the reciprocal spinor of
$\ket{\psi}$. Then
\begin{equation}
e^{-i\mathbf{F\cdot\hat{n}}\Theta}\ket{\psi}=e^{i\lambda}\ket{\psi}.
\label{eq:transformation}
\end{equation}
where $\lambda$ is given by Eq.~(\ref{Eq:lambda}).

To prove this, it is convenient to choose (without loss of generality) the
axis of rotation to be parallel to the $z$-axis: $\hat{\bf n}= \hat{\bf z}$.
Then by comparing the components of the spinors on
the left- and righthand side of Eq.~(\ref{eq:transformation}) we
see that
\begin{equation}
e^{-i \alpha \Theta}A_\alpha=e^{i\lambda}A_\alpha.
\end{equation}
This equation implies that the non-zero components of the spinor must
satisfy
\begin{equation}
\lambda\equiv - \alpha \Theta \;\; (\mathrm{mod}\ 2\pi).
\label{eq:periodic}
\end{equation}

Let us compare this result to the multiplicity of the root at the north pole, 
which corresponds to $\zeta=0$. The characteristic equation is
\begin{equation}
f_{\psi}(\zeta)=\sum_{\alpha=0}^{2S} \sqrt{2S \choose \alpha}A_{S-\alpha}^*\zeta^{\alpha}=0.
\end{equation}
If the root at zero has multiplicity $r$, then the coefficients
of the first $r$ terms $\zeta^0,\dots,\zeta^{r-1}$ are all equal to zero,
and the $r+1^{\mathrm{st}}$ coefficient is nonzero. This corresponds to
the $r+1^{\mathrm{st}}$ component of the spinor being nonzero:
\begin{equation}
A_{S-r}\neq 0.
\end{equation}
Substituting $m=S-r$ into the condition for the nonzero components Eq.~(\ref{eq:periodic}) 
gives the result Eq.~(\ref{Eq:lambda}) and the proof
is complete. Note that our discussion
has been for a rotational axis aligned with the $z$-axis, but 
the result Eq.~(\ref{Eq:lambda}) has a conceptual formulation which does not depend on 
the coordinate system.

\end{document}